\documentclass[12pt,letterpaper]{article}
\usepackage[margin=1.2in]{geometry}
\usepackage{setspace}

\usepackage{epsfig, amsthm, amsmath, amssymb, latexsym, xpatch, xcolor, mathtools, verbatim}
\usepackage[titletoc]{appendix} 
\usepackage[normalem]{ulem} % strike out % for better underline, use \uline{...}
\usepackage{tabularx}
\usepackage{stmaryrd} % for linguistics symbols
\usepackage[margin=1cm]{caption}
\usepackage{float} % to make figure exactly here
\usepackage[margin=1cm, font={small,it}]{caption} % to italicize the caption
\usepackage{wrapfig}
\usepackage{multicol} % for two-column layout

\definecolor{cardinal}{rgb}{0.8, 0.0, 0.0}

\usepackage{graphicx}

\newcommand{\op}{\begin{itemize}}
\newcommand{\ed}{\end{itemize}}

\newcommand{\opp}{\begin{quote}}
\newcommand{\edd}{\end{quote}}

\newcommand{\ope}{\begin{enumerate}}
\newcommand{\ede}{\end{enumerate}}

\newcommand{\im}{\item}

\title{Convergence to the Truth}

\author{Hanti Lin \\\\University of California, Davis \\ika@ucdavis.edu
 \\[1em] Forthcoming in {\em The Blackwell Companion to Epistemology} (3e) \\Edited by Sylvan, K., Sosa, E, Dancy, J. and Steup, M. \\Wiley Blackwell
 }

\date{} %blind
\begin{document}

\maketitle

\begin{abstract} \noindent This article reviews and develops an epistemological tradition in the philosophy of science, known as convergentism, which holds that inference methods should be assessed based on their ability to converge to the truth across a range of possible scenarios. Emphasis is placed on its historical origins in the work of C. S. Peirce and its recent developments in formal epistemology and data science (including statistics and machine learning). Comparisons are made with three other traditions: (1) explanationism, which holds that theory choice should be guided by a theory's overall balance of explanatory virtues, such as simplicity and fit with data; (2) instrumentalism, which maintains that scientific inference should be driven by the goal of obtaining useful models rather than true theories; and (3) Bayesianism, which shifts the focus from all-or-nothing beliefs to degrees of belief.
\\[1em] \noindent {\em Keywords}: Scientific Inference, Truth, Convergence, Peirce, Data Science.
\end{abstract}

%\newpage
%\addcontentsline{toc}{section}{Table of Contents}
%\tableofcontents
%\newpage

\section{Introduction} 

The epistemology of scientific inference has a rich history. According to the {\em explanationist} tradition, theory choice should be guided by a theory's overall balance of explanatory virtues, such as simplicity and fit with data (Russell 1912). % explanatory depth?
The {\em instrumentalist} tradition urges, instead, that scientific inference should be driven by the goal of obtaining useful models, rather than true theories or even approximately true ones (Duhem 1906). A third tradition is {\em Bayesianism}, which features a shift of focus from all-or-nothing beliefs to degrees of belief (Bayes 1763). It may be fair to say that these traditions are the {\em big three} in contemporary epistemology of scientific inference.

In fact, there is a fourth tradition. I am tempted to call it {\em convergentism}, although it does not yet have a widely recognized name, as this tradition is nearly lost in contemporary philosophy despite its prominence in data science (including statistics and machine learning). The central idea, traceable to the work of Peirce in the late 19th century (Peirce 1994), is that the concept of {\em convergence to the truth} should play a significant role in evaluating inference methods. This idea was further developed by Reichenbach (1938) and Putnam (1965), together with more recent contributors from statistics, machine learning, and formal epistemology. That is the story I will unfold below. Toward the end, the convergentist tradition will be briefly compared with the big three---you can expect to see not just competition, but also cooperation.

%After Peirce, you will see Reichenbach's (1938 @@@) use of a convergence theorem in reply to Hume, the later Carnap's (1963 @@@) eventual embrace of a convergence axiom in his inductive logic, and Putnam's (1967 @@@) upgrade from mere convergence to modes of convergence. More recent contributions to this tradition will be examined, too; they are less known in the wider philosophical community but mostly made in statistics, machine learning, and formal epistemology. Towards the end, the convergentist tradition will be compared with the big three---you can expect to see not just competitions but also cooperations.

\section{Peirce on Enumerative Induction}

Peirce imagined a Greek tackling a certain empirical problem---testing the hypothesis that the tide would never cease to rise every half-day:
	\opp [The Greek] had seen the tide rise just often enough to suggest to him that it would rise every half-day forever, and had proposed then to make observations to test this hypothesis, had done so, and finding the predictions successful, had provisionally accepted the theory that the tide would never cease to rise every half-day, ... . (CP 7.215) \edd 
But what justifies the Greek's acceptance of the inductive hypothesis? Peirce's answer is that the inference method in use, enumerative induction, meets a nice standard:
	\opp The only justification for this would be that it is the result of a method that, persisted in, must eventually correct any error that it leads us into. (CP 7.215) \edd 
This evaluative standard requires a {\em guarantee of eventual correction of errors}. Some important Peircean elements may not be immediately apparent from those quotes. Let me make them explicit.

% Peirce did not just claim that enumerative induction meets this standard in the Greek's tide problem; he also sketched a proof of that claim: \opp For if the tide was going to skip a half-day, he must discover it, if he continued his observations long enough. (CP 7.215) \edd 

A key Peircean element is, in a sense, {\em internalist}. That is, when inference methods are evaluated, the kind of evaluation in question can, in principle, be carried out from a {\em first-person} perspective, by the very agent tackling the empirical problem in question, such as the Greek in Peirce's example. Indeed, Peirce was not interested in an inference method that happens to converge to the truth in the actual world; he employed the modality `must' to set an evaluative standard. By `must', he had in mind a guarantee from the agent's first-person perspective, one that quantifies over the possible worlds compatible with the background assumptions or beliefs that the agent {\em does not doubt} when pursuing an empirical problem. This internalist stance is central to Peirce's objection to Cartesian skepticism (CP 5.438-52) and Peirce's praise of self-controlled, rational evaluation of one's own acts, thoughts, and reasoning (CP 1.591-611, 5.333-7).

%This is why such a guarantee of convergence to the truth can ever be mathematically proved by the agent given the background assumption in context. Peirce sought such a proof not just in the above case, but also when he did a case study on statistical inference (as we will see in section @@@).  

%To be more specific, recall the evaluation process that Peirce walked us through in the above: (i) specify an empirical problem, (ii) set a standard to evaluate inference methods, and (iii) prove that a certain method meets the standard. This evaluation process can in principle be carried out by the Greek---the agent tackling the empirical problem. 

% A second Peircean element is a kind of {\em context sensitivity}. That is, the evaluation of inference methods is sensitive to the empirical problem that the agent tackles in her context of inquiry. Indeed, when Peirce switches to an empirical problem that involves statistics, the evaluative standard in use changes---in fact, it {\em has to} change, for a systemic reason that is by no means ad hoc. More on that below (@@@). 

Another Peircean element is nicely captured by a slogan from James (1896): {\em Believe truth! Shun error!} The idea is that an inference method should be evaluated on the basis of its connection to the correction of error or, better yet, the attainment of truth: 
	\opp 
	The ... warrant for [induction] is that this method, persistently applied to the problem, must in the long run produce a convergence (though irregular) to the truth. (CP 2.775)
	\edd 
Thus, Peirce's view is a combination rarely seen in today's epistemology: that inference methods should be evaluated in an {\em internalist} way that makes explicit their connections to {\em truth-finding} (cf. [{\em add cross references}]). 
%(cf. entries  ``Internalism''*** and ``Reliabilism''***).

To clarify: It is often said that Peirce defines truth as whatever the scientific method converges to; however, if that definition were correct, it would trivialize Peirce's use of convergence to the truth in epistemology. There is strong textual evidence, though, that Peirce actually embraces a realist account of truth instead (Hookway 2000, ch. 2). Anyway, my focus will be on Peirce's epistemology, separate from his account of truth.

% Speaking of truth, there is an exegetical issue. Peirce is often interpreted to be an anti-realist about the nature of truth, defining truth as whatever the scientific method, if persistently applied, will converge to. This anti-realist interpretation, if correct, is thought to trivialize Peirce's way to justify induction in terms of convergence to the truth (Laudan @@@). On the other hand, there is textual evidence that Peirce is actually a realist, holding that what counts as true is independent of one's mind and language (Hookway @@@, Misak @@@). So, is Peirce a realist or anti-realist about the nature of truth? Whatever his metaphysical view is, it is his epistemological view that will be the focus here, to be separated from his metaphysical view.

The emphasis on the long run, however, raises an obvious concern: the long run might be too long. Or, in Carnap's (1945) terms, even if there are norms that correctly govern the long run, they are unhelpful as they say nothing about what we actually care about: norms governing the short run. Peirce's followers have developed a systematic reply to Carnap, to which I now turn.

\section{The Long Run and the Short Run}

For concreteness, I will walk you through a case study on a more precisely defined empirical problem, specified by three components:
	\op 
	\im[(i)] The {\em competing hypotheses} are `Yes, all ravens are black' and `No, not all are'.
	
	\im[(ii)] Pieces of {\em evidence} are obtained by collecting ravens and observing their colors one by one. 
	
\begin{figure}[ht]
	\centering \includegraphics[width=.9\textwidth]{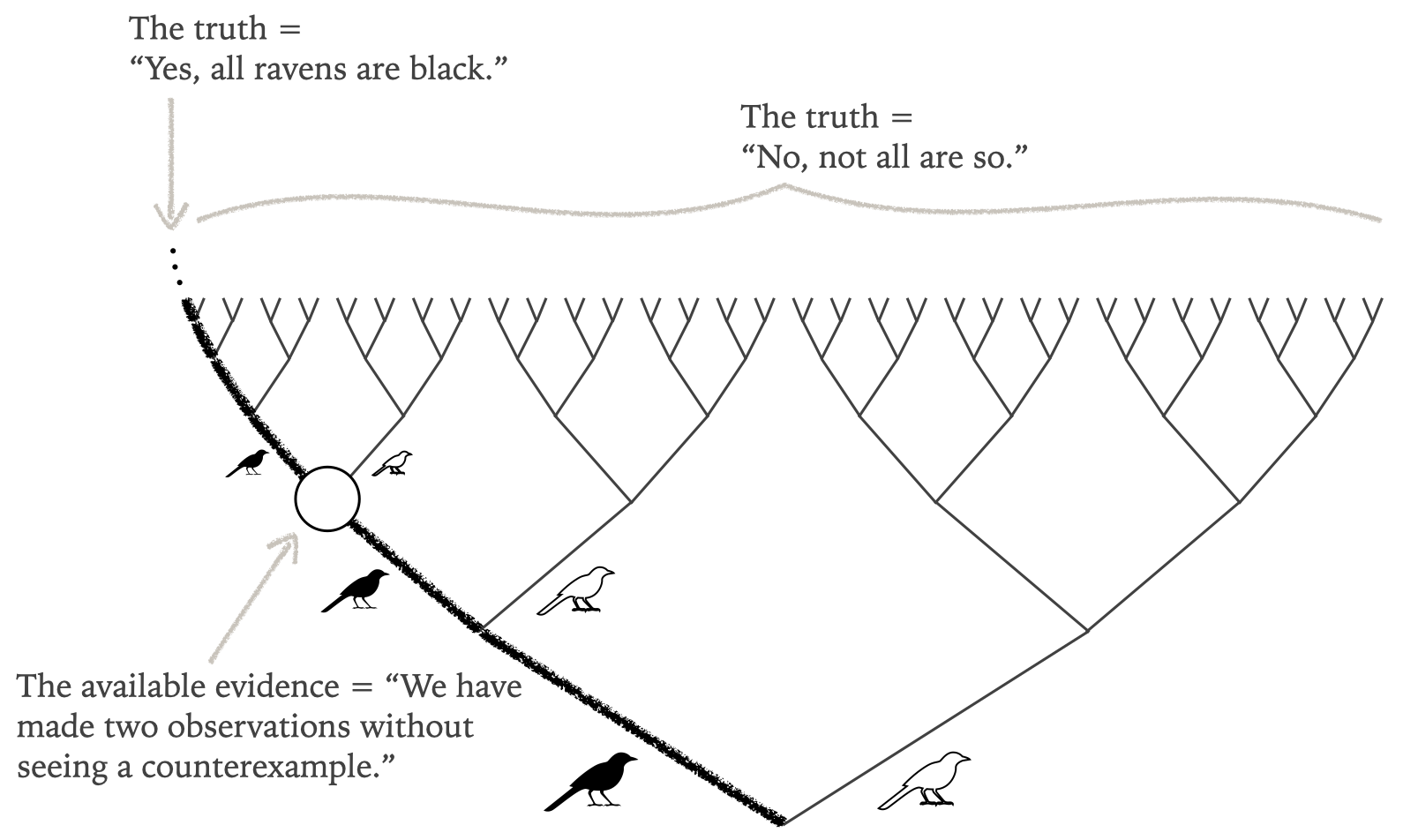}
	\caption{The raven problem is represented by a tree.}
	\label{fig-tree}
	\end{figure}

	\im[(iii)] The {\em background assumption} is that either all ravens are black or a counterexample would be observed sooner or later if the inquiry were to unfold indefinitely. 
	\ed 
Call this {\em the raven problem}. The point I want to make can be equally illustrated with a different empirical problem that alters any one of the three elements (i)-(iii), such as weakening the background assumption (Lin 2022); but then the mathematics involved would be much more complex. So, for simplicity, let me continue with the raven problem, which can be represented by the tree in Figure \ref{fig-tree}. 
The inquiry starts at the bottom, the root of the tree. Moving upward to the right signifies observing a nonblack raven (i.e. a counterexample); moving upward to the left, a black raven (or anything other than a counterexample). So, each node represents a possible body of evidence. Each branch represents a possible world, with the tip marked by the hypothesis true in that world. To clarify: although every branch is depicted as an infinite sequence, it does not represent a world in which the agent is immortal and {\em will} observe an infinite number of ravens. For example, the branch that always grows to the left only represents a world in which all ravens are black, and thus every raven observed {\em would} be black {\em if} the inquiry were to extend indefinitely. Some possible worlds are not depicted at all, as they are ruled out by the background assumption of the raven problem. 

An {\em inference method} is a function such that, whenever it receives a possible body of evidence (i.e. a node), it outputs one of the competing hypotheses or a question mark `$\texttt{?}$' to represent judgment suspension. The task at hand is to formulate some standards to evaluate inference methods.

It would be ideal if we could have an inference method that guarantees when we would obtain the truth---a guarantee of a specific amount of evidence $n$ that would yield the truth. This is a mode of convergence, which can be defined more precisely as follows:
	\opp {\bf Definition (Uniform Convergence).} An inference method $M$ for an empirical problem $P$ is said to achieve {\em uniform convergence to the truth} iff there exists an amount of evidence $n$ such that, in every possible world compatible with the background assumption of problem $P$ (i.e. in every branch of the tree), $M$ would output the truth if the number of observations were $n$ or larger. 
	\edd
This mode of convergence sets an admirably high standard---an epistemic ideal that we should strive for whenever it is achievable. Unfortunately, this standard is provably too high to be met by any inference method in the raven problem. 

\begin{figure}[ht]
	\centering \includegraphics[width=.95\textwidth]{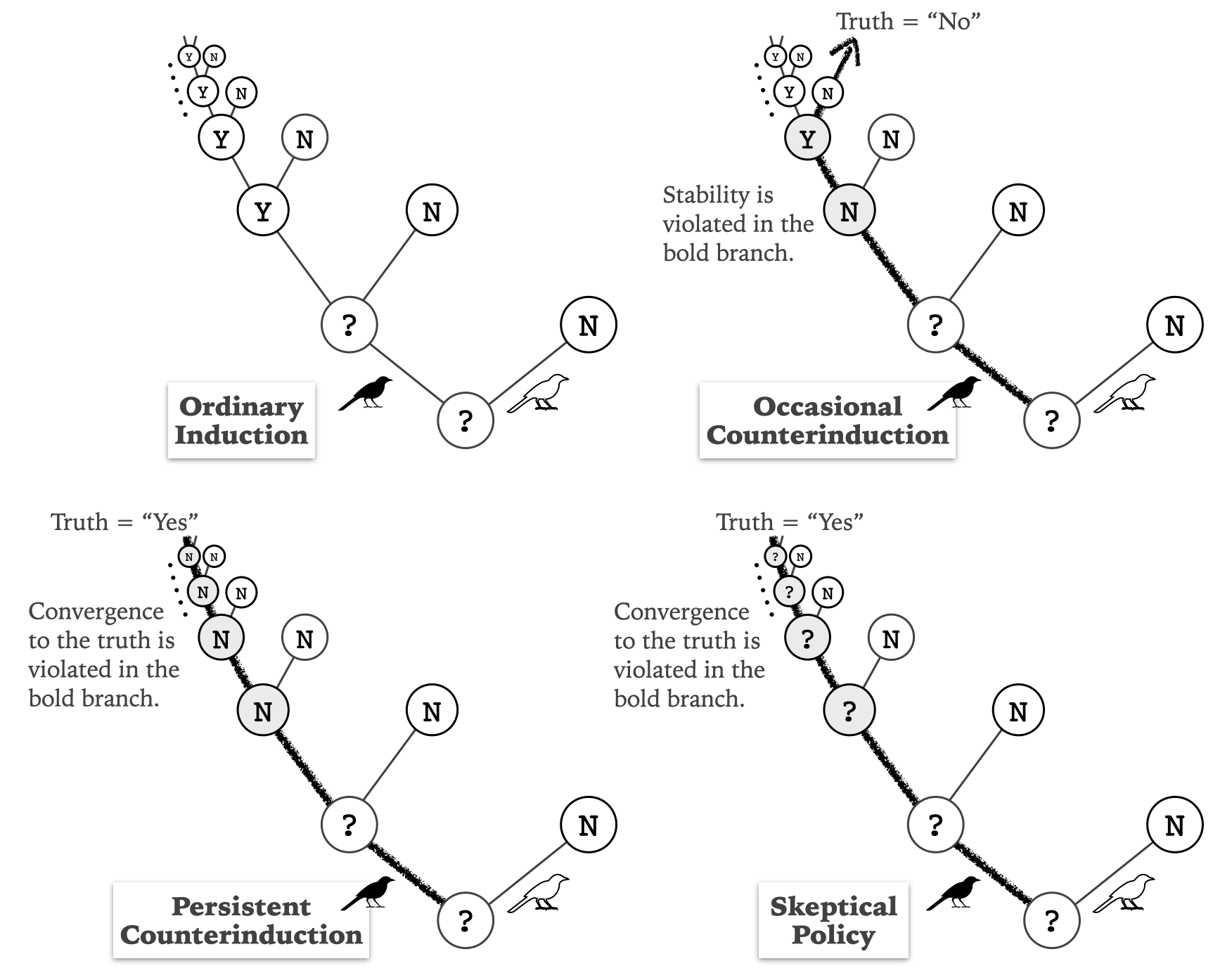}
	\caption{Four kinds of inference methods for the raven problem}
	\label{fig-stability}
	\end{figure}

A natural reaction is to try lowering the bar and {\em look for what can be achieved}. So, let's swap the two quantifiers `there exists' and `every' to define a weaker mode of convergence (and, for brevity, allow me to drop the relativity to empirical problems):
	\opp 
	{\bf Definition (Pointwise Convergence).} An inference method $M$ is said to achieve {\em pointwise convergence to the truth} iff, in every branch of the tree, there exists an amount of evidence $n$ such that $M$ would output the truth if the number of observations were $n$ or larger. 
	\edd 
The idea is that the amount of evidence needed to find the truth is allowed to vary from world to world---hence the lack of uniformity, in contrast to uniform convergence as defined above. This is exactly the mode of convergence that figures in the quotes from Peirce (whether or not he had the same motivation as presented here). This lower standard is provably achievable in the raven problem. It is achieved by, for example, the method of ordinary induction depicted in the upper left corner of Figure \ref{fig-stability}, where a `$\texttt{Y}$' denotes an output of `Yes, all ravens are black', and an `$\texttt{N}$' stands for `No, not all are'.
%%% This mode of convergence is not trivial, for it is already strong enough to rule out the skeptical policy as depicted at the lower right corner---for, in the red branch, the skeptical policy never outputs the truth. For the same reason, this mode of convergence also rules out the method of persistent counterinduction depicted at the lower left corner. 

Unfortunately, pointwise convergence only concerns the long run and thus imposes no constraint on the short run. It allows an inference method to engage in all sorts of erratic behavior (such as counterinduction) for the first few data points, before its eventual attainment of the truth. Indeed, it does not rule out the method of occasional counterinduction depicted in the upper right corner of figure \ref{fig-stability}. This is essentially Carnap's (1945) worry, originally formulated for Reichenbach's (1938) convergentist justification of induction, but it applies equally well to many convergentist justifications, including Peirce's own. 

A reply strategy began to emerge in the convergentist tradition since Putnam's (1965) seminal work. The idea is simple: there is no need to settle for a merely achievable standard such as pointwise convergence; we should, instead, {\em strive for the highest achievable}. So, let's try raising the bar by adding the following ideal to pointwise convergence: 
	\opp 
	{\bf Definition (Stability).} An inference method $M$ is said to achieve {\em stability} iff, in every branch of the tree, whenever $M$ gets the truth, $M$ would never let it go if the number of observations were to increase any further.
	\edd  
This concept simplifies Putnam's original proposal and formalizes a truth-directed ideal that Plato admires in {\em Meno}. Now, the combination of convergence and stability hits a sweet spot. It is weak enough to be achievable---achieved by the method of ordinary induction in the upper left corner of Figure \ref{fig-stability}---yet strong enough to rule out the other three inference methods in the same figure, such as counterinductions. More generally:
	\opp 
	{\bf Theorem.} In the raven problem, the combined mode of convergence to the truth plus stability is weak enough to be achievable and strong enough to rule out {\em any} inference methods that involve at least one application of counterinduction. 
	\edd 
See Figure \ref{fig-stability} for a pictorial sketch of proof. 

To recap: We have considered three modes of convergence to the truth. Listed as evaluative standards from high to low, they are:
$$\begin{array}{l}
	\textit{Uniform Convergence}
\\
	\quad\quad\quad |
\\
	\textit{Pointwise Convergence with Stability}
\\
	\quad\quad\quad |
\\
	\textit{Pointwise Convergence}
\end{array}$$
In the raven problem, the normative requirement to achieve the highest achievable standard selects the middle mode of convergence, which in turn implies a norm that governs the short run: never infer counterinductively in the raven problem---never, ever, including now. 

A short-run rabbit is thus pulled out of a long-run hat. The trick has been mostly buried in technical works, so let me try to reveal it in intuitive terms. Imagine that someone in a party is deciding whether to drink and whether to drive. She reasons as follows:
	\opp 
	I may drink, or drive, but not both. 
	\\ I must drive.
	\\ So, I must not drink.
	\edd   
This reasoning illustrates a pattern: Once a constraint is placed jointly on two things (whether-to-drink, and whether-to-drive), a constraint imposed directly on one of the two might generate a constraint on the other. Similarly, if there is a normative constraint $X$ placed jointly on two things---the long run and the short run---then a convergentist constraint on the long run might generate a nontrivial constraint on the short run. Such an $X$ must then be a {\em diachronic} constraint, and that is the trick in reply to Carnap. The mode of stable convergence defined above is indeed diachronic, for it concerns the retention of truth as evidence accumulates. Convergentists have explored other diachronic candidates for $X$, which will be presented below when needed.

\section{A Framework for Convergentism}

The above case study on the raven problem actually illustrates a framework, first adumbrated in Putnam (1965) and later practiced by Kelly (1996) and Schulte (1999) in a wider domain. A very general statement of the core thesis was articulated by Lin (2022) and I propose to make it clearer as follows:
	\opp {\bf The Core Thesis of Achievabilist Convergentism.} In any empirical problem, a necessary condition for an inference method to be justified is that it achieves the highest achievable mode of convergence to the truth---pending a specification of the correct hierarchy of modes of convergence as evaluative standards.
	\edd 
This thesis sets up what may be called the {\em achievabilist} framework for convergentism, for lack of a standard name. This framework encourages the exploration of modes of convergence, such as the mode of stable convergence, which is used to address Carnap's worry. 

This framework also features a kind of context-sensitivity: the appropriate standard for assessing inference methods should be the highest achievable, which is sensitive to a contextual factor: the empirical problem that one tackles in one's context of inquiry. In fact, when one switches to the context of a statistical problem, the epistemic standards presented above all become unachievable, which requires convergentists to explore lower standards---weaker modes of convergence. To illustrate how that may be done, let's think about statistics, which brings us back to Peirce.

\section{Peirce on Statistics}\label{sec-stat}

Peirce once studied a classic problem in statistics. An urn contains an unknown number of black and white balls. We ask: What is the true proportion of white balls in the urn? The three components of this problem are as follows:
	\op 
	\im The {\em competing hypotheses} are the rational numbers in the unit interval (i.e. the possible proportions). 
	\im {\em Evidence} is to be collected by randomly drawing balls with replacement. 
	\im The {\em background assumption} is that different draws are probabilistically independent. 
	\ed 
Call this the {\em white ball problem}. To evaluate inference methods for this problem, Peirce proposed to employ the following mode of convergence, where, by probability, he meant physical chance:
	\opp 
	{\bf Definition (Statistical Consistency).} An inference method $M$ for an empirical problem is said to achieve {\em statistical consistency} iff (i) in any possible world compatible with the background assumption, there exists $n$ such that inference method $M$ would highly probably produce a guess that gets close to the truth if the sample size were $n$ or larger, and (ii) inference method $M$ has the above property for any threshold of high probability less than 1 and for any nonzero threshold of closeness.
	\edd 
Peirce studied this mode of convergence as an evaluative standard in an 1878 paper titled ``The Probability of Induction'' (CP 2.669-93). It is unclear whether Peirce influenced any statisticians of his time, but this stochastic mode of convergence was popularized by statistician Fisher (1925: sec. I.3). It is now generally regarded in frequentist statistics as a minimum qualification for any justified statistical methods, with different versions for different statistical problems, including estimation, hypothesis testing, and regression (i.e. curve fitting).

Peirce never explained why he employed different evaluative standards in two different empirical problems, the white ball problem and the Greek's tide problem (which is equivalent to the raven problem). From an achievabilist hindsight, he had to employ different standards. The Greek's tide problem is easy enough to allow for a guarantee of getting exactly the true answer at least when the amount of evidence is arbitrarily large. But this standard is too high to be achievable in the white ball problem. So, let's try lowering the bar for that problem: ``getting exactly the truth'' can be downgraded to ``{\em highly probably} getting exactly the truth'', which can be further downgraded to ``highly probably getting {\em close} to the truth''. This line of thought motivates statistical consistency as a lower standard, and Peirce did show how it can be achieved in the white ball problem (by using the law of large numbers).

Carnap's worry still needs to be addressed for statistical problems. The diachronic trick presented above still applies; an example will be provided in a broader context when I wrap up.

\section{Closing: Comparison with the Big Three}\label{sec-closing}

How does the convergentist tradition fare against the big three mentioned in the introduction? 

First of all, the Bayesian tradition need not be a rival to convergentism. An alliance has been proposed in statistics as a partial solution to Bayesians' perennial problem of the priors---the problem of identifying the correct norms that govern prior credences (see [{\em add cross references}]). Here is a convergentist constraint on priors due to statistician Freedman (1963):
	\opp {\bf Definition (Bayesian Consistency).} A prior is said to achieve {\em Bayesian consistency} in an empirical problem iff it is guaranteed (under just the background assumption of that problem) that this prior, when guided by the diachronic rule of conditionalization, would have a high chance of leading to posterior credences that converge to the truth among the considered hypotheses if the amount of evidence were to accumulate indefinitely---for any threshold of high chance less than $1$. 
	\edd 
%%% Like any other mode of convergence to the truth, this is sensitive to the empirical problem in context (for the exact contents of `guaranteed' and `competing hypotheses' are context-sensitive). 
Freedman did not recognize that this is another implementation of the diachronic trick that replies to Carnap: a combination of long-run convergence with a diachronic constraint, which in this case is conditionalization. The resulting constraint on the priors---and hence on the short run---turns out to be surprisingly strong in some interesting empirical problems, stronger than what traditional Bayesians have to offer. In particular, Bayesian consistency implies a version of Ockham's razor in statistical problems of curve-fitting (Diaconis \& Freedman 1998). Similarly, a chance-free counterpart of Bayesian consistency implies another version of Ockham's razor in problems of testing deterministic hypotheses (Lin 2022). It remains to be seen whether convergentist Bayesianism is better than the more traditional varieties of Bayesianism.

The explanationist tradition can benefit from convergentism, too. According to explanationism, theory choice should be based on a theory's overall balance of explanatory virtues, so it should be based on Ockham's razor if simplicity is among these virtues. But which version of Ockham's razor? That is, which particular trade-off between simplicity and other virtues such as fit with data? Under which conception of simplicity? And under which conception of fit? Explanationists often think that the choice is to be justified by appeal to intuition (Swinburne 1997). However, in complex empirical problems, working scientists often find that they lack the intuition needed to justify one version of Ockham's razor over another---and this is where convergentists can assist. The previous paragraph already referred to two examples of convergentist justifications of particular versions of Ockham's razor, in curve-fitting problems and in problems of testing deterministic hypotheses. Let me add one more example: in problems of testing statistical hypotheses, Genin (2018) justifies another version of Ockham's razor by combining a mode of convergence with a stochastic version of stability (which he calls {\em progressiveness}, a guarantee that the chance of finding the truth would never drop too much if the sample size were increased by any arbitrary finite amount).

Now, let's turn to the instrumentalist tradition. Due to their emphasis on the pursuit of usefulness instead of truth, instrumentalists might appear to have to dismiss the significance of convergence to the {\em truth}. However, this is merely an appearance, I submit. The usefulness of a scientific model is often taken to include at least predictive accuracy, but predictive accuracy is a contingent matter, depending on what the actual world is like. Thus, choosing a useful model is no trivial task. This issue is addressed in a foundational branch of machine learning, known as {\em statistical learning theory} (Shalev-Shwartz \& Ben-David 2014, part I). It assesses learning algorithms based on some modes of convergence, such as a guarantee to have a high chance of converging to the most useful of the predictive devices under consideration, where usefulness is identified with the chance of accurate prediction. 
	To be sure, this does not sound like convergence to the truth. But let's be careful: convergence to the most useful one is still convergence to a certain truth, the true answer to this practical question: {\em Which one is the most useful?} Instrumentalists can find a home in machine learning---a convergentist home.

The comparisons just made are quite rudimentary due to the lack of established literature. Much work is still needed to adjudicate the competition between convergentism and the other three traditions, as well as to explore opportunities for their cooperation.

%\end{document}
\section*{References}

\begin{description}
\im Bayes, T. (1763) ``An Essay towards Solving a Problem in the Doctrine of Chances'', {\em Philosophical Transactions of the Royal Society}: 330-418. Reprinted with Biographical Note by Barnard, G.A., in (1958) {\em Biometrika} 45: 293-315.

\im Carnap, R. (1945) ``On Inductive Logic'', {\em Philosophy of Science}, 12(2): 72-97.

\im Diaconis, P. \& Freedman, D. (1998) ``Consistency of Bayes Estimates for Nonparametric Regression: Normal Theory'', {\em Bernoulli}, 4(4): 411-444.

\im Duhem, P. (1906/1954) {\em The Aim and Structure of Physical Theory}, Princeton University Press.

\im Fisher, R.A. (1925) {\em Statistical Methods for Research Workers}, Oliver \& Boyd.

\im Freedman, D. (1963) ``On the Asymptotic Behavior of Bayes' Estimates in the Discrete Case'', {\em The Annals of Mathematical Statistics}, 34(4): 1386-1403.

\im Genin, K. (2018) {\em The Topology of Statistical Inquiry}, PhD Dissertation, Carnegie Mellon University.

%\im Gold, E.M. (1967) ``Language Identification in the Limit'', {\em Information and Control}, 10(5): 447-474.

%\im Good, I.J. (1971) ``46656 Varieties of Bayesians'', {\em The American Statistician}, 25: 62-63.

\im Hookway, C. (2000) {\em Truth, Rationality, and Pragmatism: Themes from Peirce}, Oxford University Press.

\im James, W. (1896) ``The Will to Believe'', {\em The New World}, 5: 327-347.

\im Kelly, K.T. (1996) {\em The Logic of Reliable Inquiry}, Oxford University Press.

\im Lin, H. (2022) ``Modes of Convergence to the Truth: Steps toward a Better Epistemology of Induction'', {\em The Review of Symbolic Logic}, 15(2), 277-310.

%%% \im Migotti, M. (1998) ``Peirce's Double-aspect Theory of Truth'', {\em The Canadian Journal of Philosophy}, Suppl. 24: 75-108.

%\im Mohri, M., Rostamizadeh, A., \& Talwalkar, A. (2018) ``Foundations of Machine Learning'', MIT Press.

\im Peirce, C.S. (1994) {\em Collected Papers of Charles Sanders Peirce} (Volumes I-VIII), InteLex Corp.

\im Putnam, H. (1965) ``Trial and Error Predicates and a Solution to a Problem of Mostowski'', {\em The Journal of Symbolic Logic}, 30(1): 49-57.

\im Reichenbach, H. (1938) {\em Experience and Prediction: An Analysis of the Foundation and the Structure of Knowledge}, University of Chicago Press.

\im Russell, B. (1912) {\em The Problems of Philosophy}, Williams and Norgate.

\im Schulte, O. (1999) ``Means-Ends Epistemology'', {\em The British Journal for the Philosophy of Science} 79(1), 1-32.

\im Shalev-Shwartz, S. \& Ben-David, S.  (2014) {\em Understanding Machine Learning: From Theory to Algorithms}, Cambridge University Press.

%\im Shao, J. (1997) ``An Asymptotic Theory for Linear Model Selection'', {\em Statistica Sinica}, 221-242.

\im Swinburne, R. (1997) {\em Simplicity as Evidence of Truth}, Marquette University Press.
\end{description}

\end{document}